\documentclass[apj]{emulateapj}
\usepackage[dvipsnames, usenames]{color}
\usepackage{latexsym}
\usepackage{amsmath}
\usepackage{graphics}

\shorttitle{The Origin of the EUV Late Phase}
\shortauthors{Hock et al.}

\begin{document}

    \title{The Origin of the EUV Late Phase: A Case Study of the C8.8 Flare on 2010 May 5}
    \author{R.~A. Hock,  T.~N. Woods}
    \affil{Laboratory for Atmospheric and Space Physics, University of Colorado, Boulder, CO 80303, USA}

    \author{J.~A. Klimchuk}
    \affil{NASA Goddard Space Flight Center, Solar Physics Laboratory, Greenbelt, MD 20771, USA}

    \author{F.~G. Eparvier, and A.~R. Jones}
    \affil{Laboratory for Atmospheric and Space Physics, University of Colorado, Boulder, CO 80303, USA}

    \email{rachel.hock@lasp.colorado.edu}

% ===============================
% Abstract
% ===============================

    \begin{abstract}
    Since the launch of NASA's Solar Dynamics Observatory on 2010 February 11, the Extreme ultraviolet Variability Experiment (EVE) has observed numerous flares.  One interesting feature observed by EVE is that a subset of flares exhibit an additional enhancement of the 2-3 million K emission several hours after the flareÕs soft X-ray emission.  From the Atmospheric Imaging Assembly (AIA) images, we observe that this secondary emission, dubbed the EUV late phase, occurs in the same active region as the flare but not in the same coronal loops. Here, we examine the C8.8 flare that occurred on 2010 May 5 as a case study of EUV late phase flares. In addition to presenting detailed observations from both AIA and EVE, we develop a physical model of this flare and test it using the Enthalpy Based Thermal Evolution of Loops (EBTEL) model.
    \end{abstract}
    \keywords{Sun: activity --- Sun: flares --- Sun: UV radiation}

% ===============================
% Section 1: Introduction
% ===============================

    \section{Introduction}
    \label{sec:intro}
    X-ray and extreme ultraviolet (EUV) emission from the solar corona has
long been used to provide insight into the dynamics and evolution of
flares. Satellites such as {\it Hinode} and the {\it Solar and
Heliospheric Observatory} ({\it SOHO}) include imaging spectrographs
with high spectral resolution but limited spatial and wavelength
coverage, as well as broadband X-ray and EUV imagers. The EUV
Variability Experiment (EVE) onboard NASA's {\it Solar Dynamics
Observatory} ({\it SDO}) adds another tool for flare studies.  EVE
measures the solar spectral irradiance (disk-integrated radiance) in
the EUV and overlaps spectrally and temporally with the Atmospheric
Imaging Assembly (AIA), which takes high-resolution, full-disk solar
images at seven EUV bands with a twelve-second cadence. Even though
EVE measures the full-disk radiative output of the Sun, the low
levels of solar activity at the beginning of the {\it SDO} mission
allow for the study of individual flares with little or no
contamination from other events. The wavelength range (6-105 nm),
spectral resolution (0.1 nm) and continuous time coverage of EVE
provides observations at a wide range of temperatures (7,000 K to 10
MK), capturing the full flare and post-flare evolution
of the transition region and corona.

    Several of the first flares observed by {\it SDO} in May 2010 had an unexpected irradiance signature: EVE observes two distinct peaks in EUV emission at about 2 MK \citep{woods_euv_late_phase}.  The first enhancement peaks minutes after the flare's soft X-ray emission peaks and is part of the gradual phase of the flare while the second, dubbed the ``EUV late phase,'' occurs several hours after the flare's soft X-ray emission. From AIA images, we can tell that these EUV late phase emissions occur in the same active region as the flare but not in the same coronal loops as the gradual phase.  \citet{woods_euv_late_phase} introduces some of the EVE flare observations and the EUV late phase, and here, we examine the C8.8 EUV late phase flare that occurred on 2010 May 5 as a detailed case study of an EUV late phase flare. Our goal is to present detailed observations  from both EVE  and AIA as well as model the radiative output of the flare using the Enthalpy Based Thermal Evolution of Loops (EBTEL) model developed by \citet{klimchuk_ebtel}.  In order to aid in the understanding of this complicated event, we start with physical picture of the flare in Section \ref{sec:model}. In Section~\ref{sec:observations} we present the detailed observations from SDO, and finally in Section ~\ref{sec:ebtel} conclude with results of the EBTEL modeling.

% ===============================
% Section 2: Cartoon
% ===============================

    \section{Physical Model}
    \label{sec:model}

\begin{figure}[htbp]
    \begin{center}
    \includegraphics[width=0.7\columnwidth]{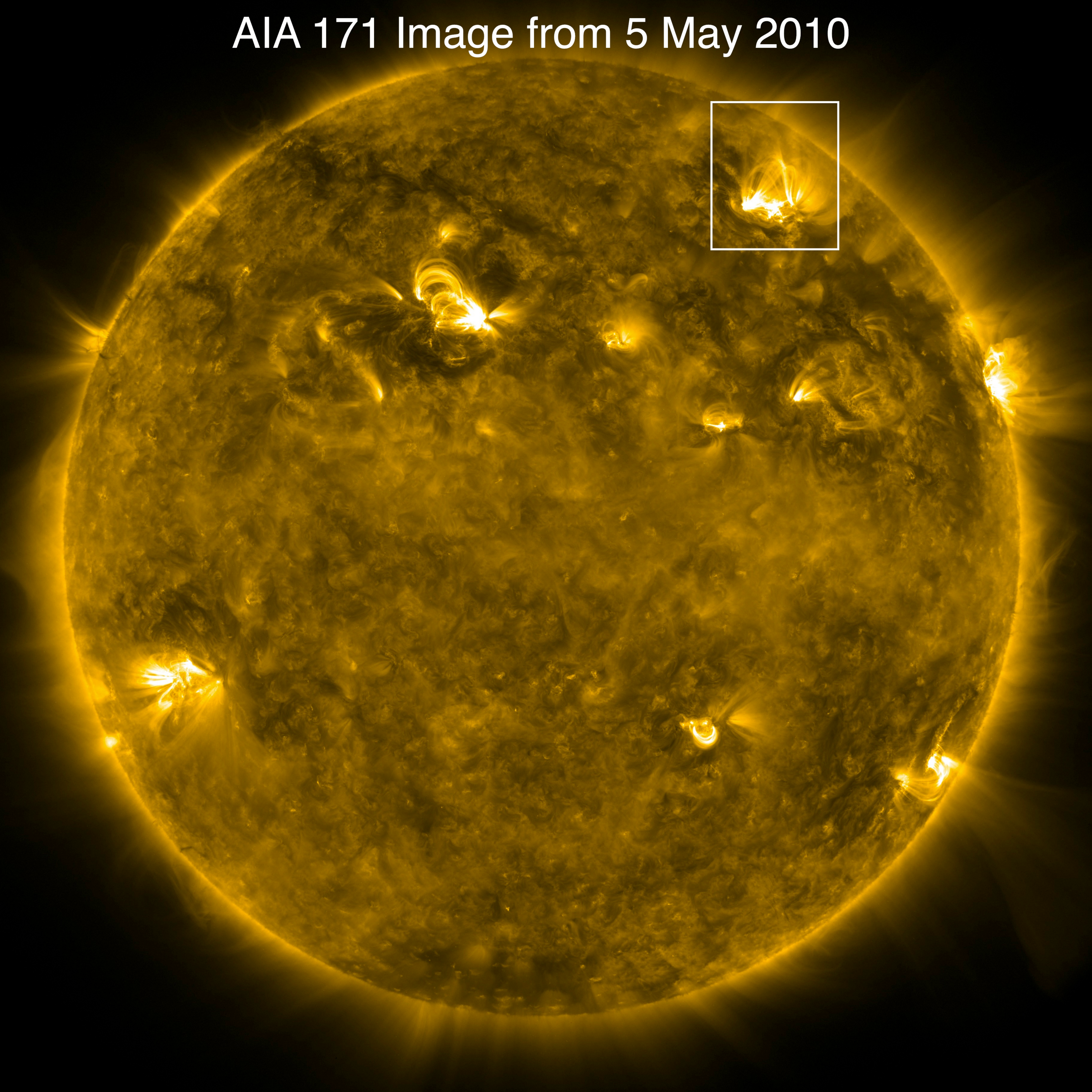}
    \end{center}
    \caption{Image from AIA on 5 May 2010 showing the location on AR 11069 near the northwest limb.}
    \label{fig:context_image}
    \end{figure}

    To provide a context for the observations and modeling presented in the following sections, we begin with an overview of the flare.  The C8.8 flare on 2010 May 5 occurred in NOAA Active Region 11069 near the northwest limb (Figure~\ref{fig:context_image}). From 2010 May 1 to 2010 May 8, this active region produced eight flares larger than C1.0, six of which had an EUV late phase.

    % Figure: Context image

    The EUV late phase is a secondary peak in EUV irradiance lightcurves for 2-3 MK coronal emissions with no corresponding emission in the GOES soft X-ray flux (SXR) or hotter emission lines such as Fe {\sc xx}/{\sc xxiii}. From \citet{woods_euv_late_phase}, there are four criteria for the EUV late phase: 
    (1) a second peak in the lightcurves of the Fe {\sc xv} and Fe {\sc xvi} emissions after the GOES X-ray peak is observed by EVE; 
    (2) there are no corresponding enhancements in GOES soft X-ray $0.1-0.8$ nm  (SXR) or hotter emission lines; 
    (3) the flare is associated with an eruption seen in EUV images; 
    (4) the secondary emission comes from a set of longer (and higher) loops than the original flaring loops.

    % Figure: SXR and EVE lightcurves to define phase of flare
    \begin{figure*}
    \begin{center}
    \includegraphics[width=\textwidth]{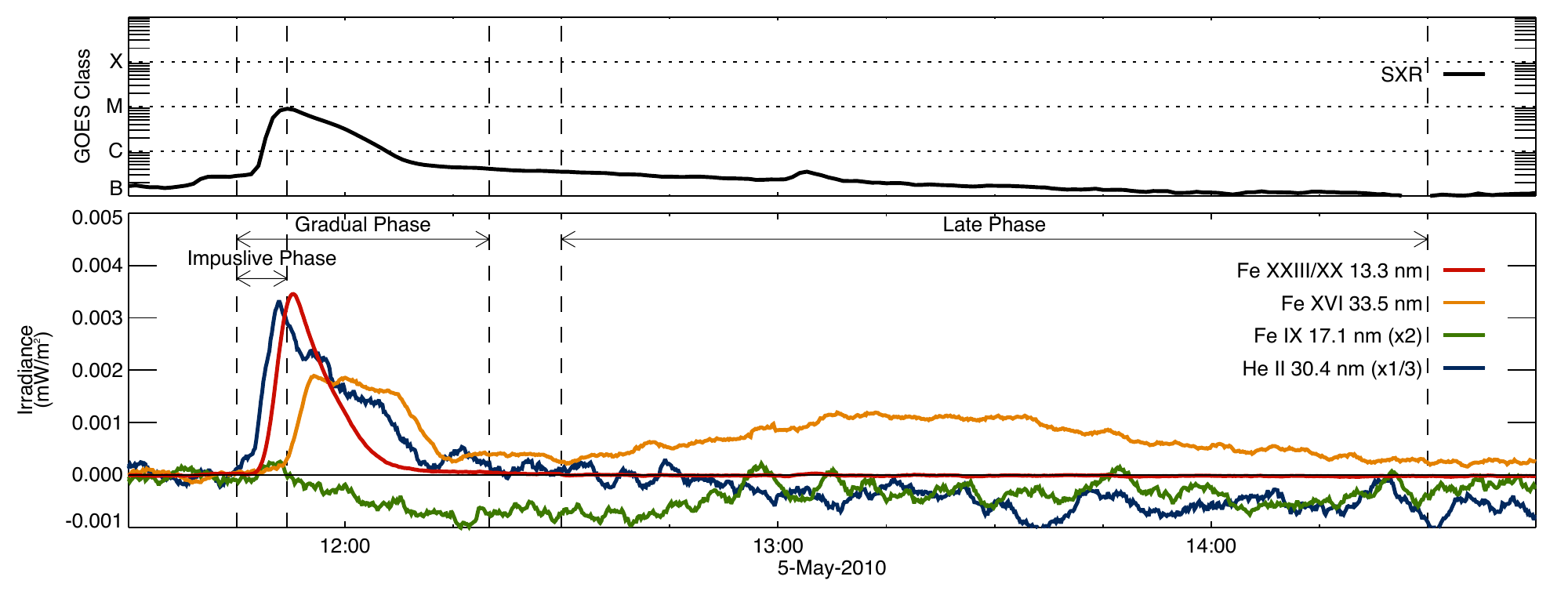}
    \end{center}
    \caption{The GOES soft X-ray flux and highlighted EVE lines show the three distinct phases of the C8.8 flare on 2010 May 5.  The impulsive phase (11:45-11:52 UT) is characterized by the increase in the transition region He {\sc ii} line; the gradual phase (11:52-12:20 UT) is observed in both hot and warm (2-10 MK) coronal lines while the EUV late phase (12:30-14:30 UT) is only observed in the warm (2-3 MK) lines. The irradiance shown has had the pre-flare level subtracted.}
    \label{fig:sxr_eve_flare_phases}
    \end{figure*}

      The C8.8 flare on 2010 May 5 (GOES peak at 11:52 UT) is one of the clearest examples of an EUV late phase flare observed by{\it SDO} during the first year of normal science operations (2010 May 1 to 2011 April 30).  The first two criteria (second peak in Fe {\sc xvi} and no corresponding emission in GOES SXR) for the EUV late phase are clearly seen in Figure~\ref{fig:sxr_eve_flare_phases}.  Figure~\ref{fig:sxr_eve_flare_phases}a shows the GOES SXR lightcurve for this flare while Figure~\ref{fig:sxr_eve_flare_phases}b shows the irradiance lightcurves from EVE for a hot coronal line (Fe {\sc xx}/{\sc xxiii} at 13.3 nm), warm coronal line (Fe {\sc xvi} at 33.5 nm), cool coronal line (Fe {\sc ix} at 17.1 nm), and a transition region line (He {\sc ii} at 30.4 nm).

      Using the lightcurves from these emission lines, we can identify four key features of this flare. First, He {\sc ii} peaks prior to the GOES SXR. This is the impulsive phase of the flare, generally seen in hard X-rays and chromospheric and transition region emission lines. Second, both Fe {\sc xx}/{\sc xxiii} and Fe {\sc xvi} peak shortly after the GOES SXR and form the gradual phase of the flare. Third, Fe {\sc ix}, a cool coronal emission, actually shows a slight decrease during the flare.  This coronal dimming is the result of mass loss due to a coronal mass ejection (CME).  Evidence for a CME associated with the 2010 May 5 C8.8 flare is discussed in Section~\ref{sec:eruption_breakout}.  Finally, there is a second peak seen in Fe {\sc xvi} (2-3 MK) but not in GOES SXR or the Fe {\sc xx}/{\sc xxiii} line. This enhancement, the EUV late phase, starts about thirty minutes after the peak in GOES SXR and slowly increases over an hour before gradually decaying to the pre-flare levels after two hours.

\begin{figure}[htbp]
    \begin{center}
    \includegraphics[width=0.7\columnwidth]{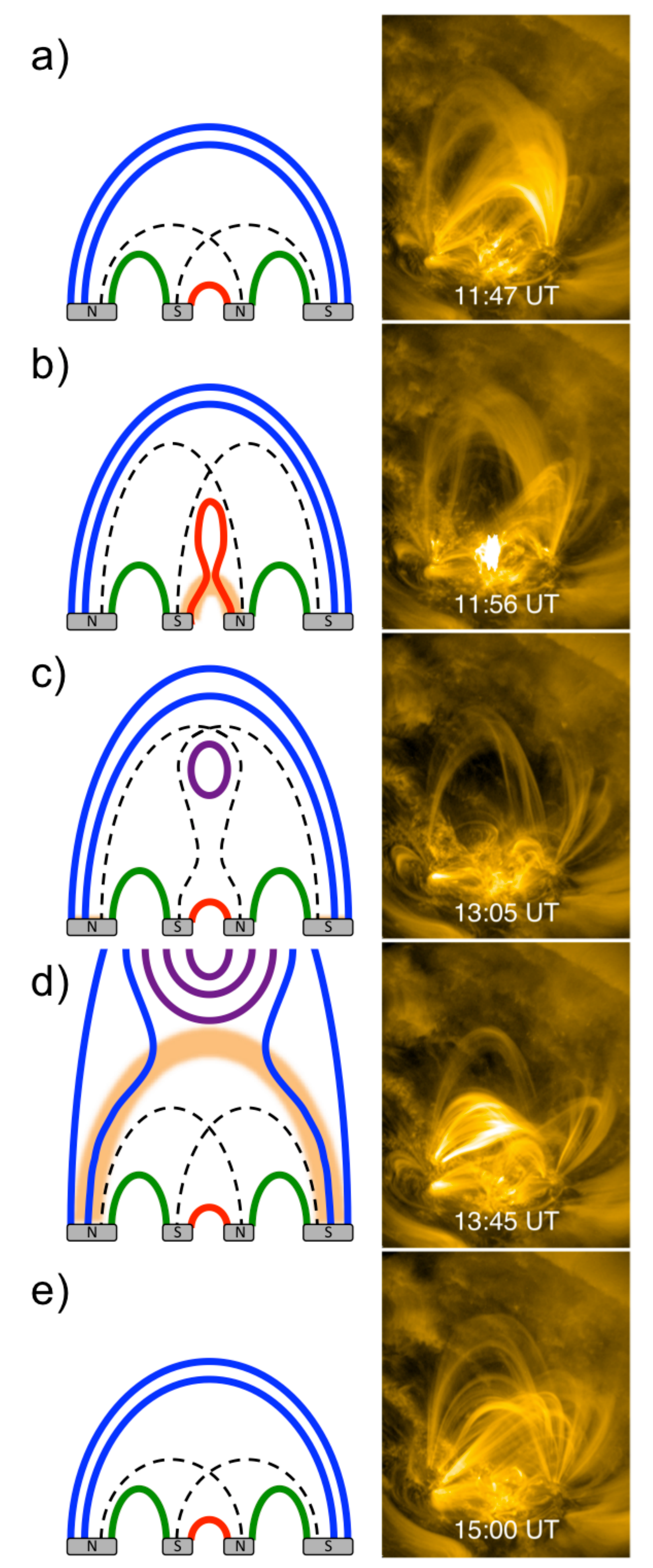}
    \end{center}
    \caption{Schematic cartoon of the 2010 May 5 C8.8 flare (left) and images from AIA 171 (right) showing (a) the pre-flare configuration, (b) the main phase of the flare, (c) eruption and formation of CME (purple), (d) the EUV late phase, and (e) the recovery of the active region to a state similar to the pre-flare configuration.   The separatrices marking the boundary between the inner (red) and outer loops (blue) are denoted by the dashed lines.  The side lobes (green) remain unchanged during the flare. The orange regions show where and when we expect the brighter emission based on what loops are undergoing reconnection.}
    \label{fig:cartoon}
    \end{figure}

    Spatially resolved observations from AIA allow us to verify
the third and fourth EUV late phase criteria (eruption and overlying
loops) and to develop a physical picture for the evolution of this
flare. In this purely schematic model (Figure~\ref{fig:cartoon}),
the main phase, which includes the gradual and impulsive phases, and
the EUV late phase are connected through the breakout model of CME
initiation \citep{antiochos_breakout_reconnection2}.

    The active region starts in a classic quadrupolar configuration
(Figure~\ref{fig:cartoon}a). In a quadrupolar region, there are four
flux systems: inner loops (red) defining the core of the region, two
side lobes (green), and outer loops (blue) overlying the inner and
side loops. These flux systems are separated by two separatrices
(dashed lines), which intersect at a null point above the inner
loops. The separatrices are replaced by a separatrix surface and
spine in a fully 3D configuration.  In the breakout picture,
eruption is initiated by reconnection at the null point, which
converts red and blue field lines from the inner and outer loops
into green field lines of the side loops.  This releases some of the
magnetic tension and allows the inner flux system to expand upward.
Shortly after the process begins, internal reconnection within the
inner flux system pinches off a plasmoid (flux rope in 3D) and the
eruption accelerates (Figure~\ref{fig:cartoon}b and c). This is
known as flare reconnection and is much more energetic than the
breakout reconnection that occurs at the rising null point. The
flare reconnection produces an arcade of compact loops, often with a
cusp-shaped top.  Emission from these loops is responsible for the
gradual phase emission and peaks at successively cooler temperatures
as the loops cool.

As the eruption and flare proceed, enough of the inner (red) flux
system may reconnect to allow internal reconnection of the outer
(blue) flux system (Figure~\ref{fig:cartoon}d).  The process is
similar to flare reconnection, but is less energetic due to the
weaker magnetic fields that are involved.  The newly created loops
are longer, higher, and cooler than those of the compact flare
arcade. They span the active region.  It is these loops that produce
most of the EUV late phase emission.  Because they cool as they retract
(the reconnection occurs at much greater heights than shown in the
cartoon), the hotter loops tend to overlie the cooler ones.  The
active region is eventually returned to something resembling its
initial state, though with less stress (shear) and less magnetic
energy.

We note a second possible origin for the the long (blue) loops in
Figure~\ref{fig:cartoon}e.  Breakout reconnection may continue above
the plasmoid long after the eruption is initiated.  This
reconnection is between the oppositely directed purple and blue field
lines in Figure~\ref{fig:cartoon}c.  It creates new field lines that
no longer arch over the plasmoid and are therefore free to retract
to the active region below.  Which scenario dominates is unclear at
the present time.  A possible way to distinguish between them would
be via {\it in situ} measurements of the poloidal magnetic field of
the plasmoid (flux rope).  The poloidal field should have opposite
sign in the inner and outer parts of the plasmoid in the first
scenario, but like sign in the second scenario.

    % FIgure: Flare Cartoon

    In the following section, both the EVE and AIA observations are examined in more detail to further support this model of this EUV late phase flare.

% ===============================
% Section 3: Observations
% ===============================

    \section{Observations}
    \label{sec:observations}

    The strength of {\it SDO} in studying solar flares arises from coupling broadband EUV images observed by AIA with the EUV spectral irradiance measured by EVE.  The data from EVE used in this paper is from the Multiple EUV Grating Spectrographs (MEGS) that provide EUV irradiance spectra with a spectral resolution of 0.1 nm from 6 to 105 nm, a high time cadence of 10 seconds, and accuracy better than 20\%. \citet{woods_eve_overview} provide an overview of EVEÕs science plans, instrument design, and data products while \citet{hock_megs_cal} provide an overview of the MEGS instrument and its calibration.

\begin{figure*}[p!]
    \begin{center}
    \includegraphics[width=\textwidth]{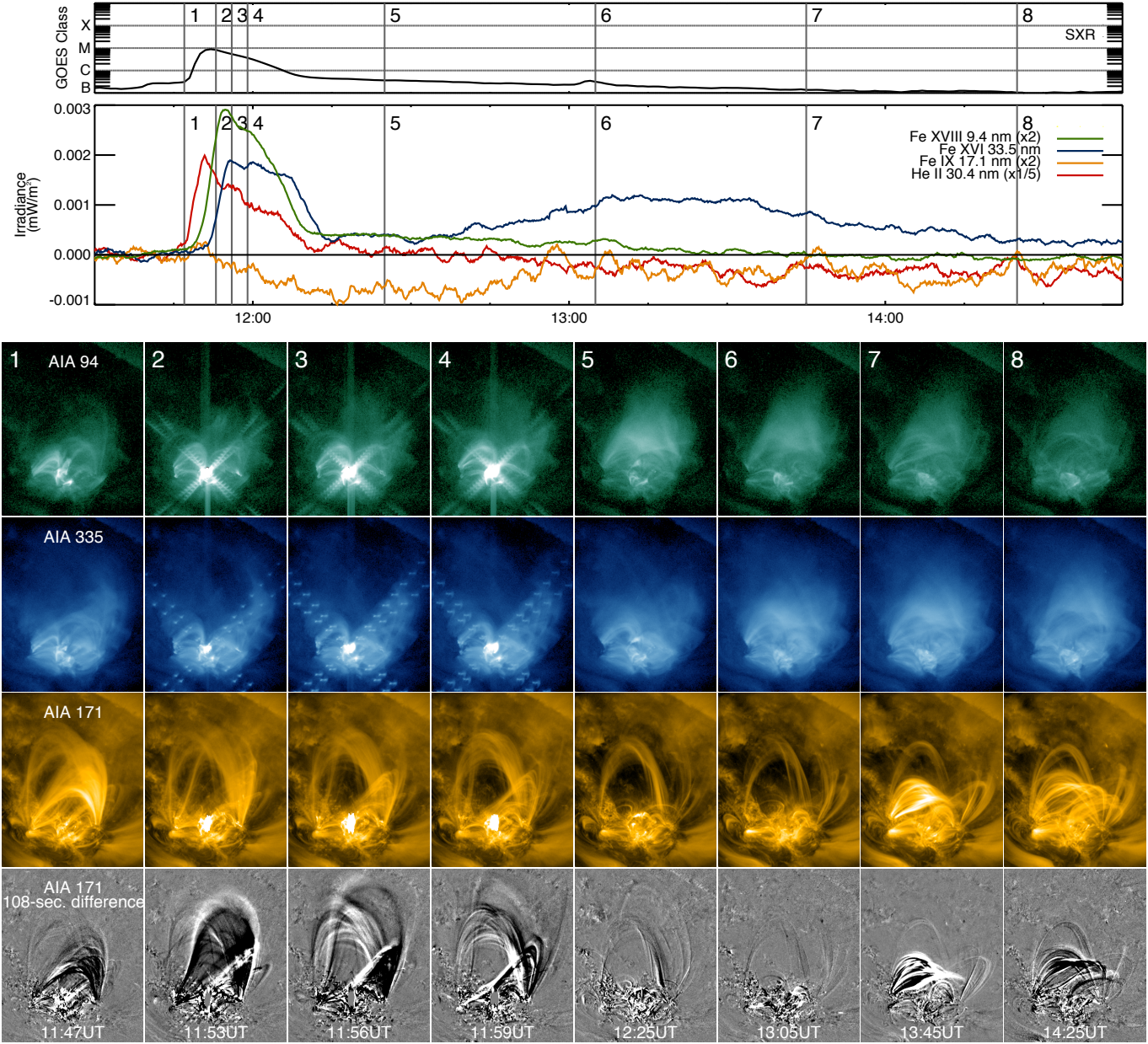}
    \end{center}
    \caption{Images from AIA 94, AIA 335, and AIA 171 channels as well as AIA 171 108-second difference images show the evolution of the active region before, during, and after the flare with time moving from left to right.  The difference images highlight the movement of loops/mass as the color transitions from black to white. The time of each image sequence is designated by a number and shown in GOES SXR and EVE lightcurves (top).  The EVE lightcurves include the major line in each of the AIA channels shown as well as He II 30.4 nm. }
    \label{fig:aia_eve}
    \end{figure*}

    For temporally isolated flares like the C8.8 flare on 2010 May 5, we can determine the time-dependence of the flare spectra over a wide range of temperatures by subtracting a pre-flare spectrum. For this event, the pre-flare spectrum is the average of observations from 11:30-11:45 UT (90 integrations), corresponding to the minimum in the GOES SXR flux prior to the flare. Figures~\ref{fig:sxr_eve_flare_phases}, \ref{fig:aia_eve}, \ref{fig:eve_impulsive_phase}, and \ref{fig:eve_gradual_phase} show the time evolution of the GOES SXR flux and select EUV lines measured by EVE during the flare. Note that the irradiances shown have had the pre-flare level subtracted.

    Table~\ref{table:eve_lines} includes the ion, wavelength, and peak formation temperature of all EVE emissions discussed in this paper.  Equilibrium ionization is assumed, though see \citet{bk2011}. The EVE lines cover a temperature range of 50,000 K (He {\sc ii}) to 10 MK (Fe {\sc xxiii}/{\sc xx}). Where applicable, the appropriate AIA channel is also identified. Note that the EVE emission is at 0.1 nm resolution; whereas, the AIA channels are over broader bands and thus have more line blends than the EVE measurements.  While EVE measures the spectral irradiance with a 10-second integration, to reduce noise, a three-minute (18 integrations) boxcar smoothing function is applied to the data that are used throughout this paper.

    % Table: EVE lines
    \begin{table}[htdp]
    \caption{Emission lines from EVE}
    \begin{center}
    \begin{tabular}{crrc}
    Ion & Wavelength (nm) & log T (K) & AIA Channel\\
    \hline
    Fe {\sc xxiii}/Fe {\sc xx} & 13.285 & 6.97 & A131\\
    Fe {\sc xviii} & 9.393 & 6.81 & A94\\
    Fe {\sc xvi} & 33.541& 6.43 & A335\\
    Fe {\sc xv} & 28.415 & 6.30&\\
    Fe {\sc ix} & 17.107 & 5.81& A171\\
    He {\sc ii} & 30.378 & 4.70 & A304 \\
    He {\sc i} & 58.433 & 4.16 &
    \end{tabular}
    \end{center}
    \label{table:eve_lines}
    \end{table}

    Lightcurves from EVE are used to identify the EUV late phase but in order to understand the origin and nature of the EUV late phase, spatially resolved observations, such as the EUV images from AIA, must be examined. AIA has four telescopes, each with a 4096$\times$4096 pixel CCD. These provide full-disk solar images in seven EUV bands and three UV/visible bands. AIA images have a spatial resolution of 0.6 arc-seconds and a nominal time cadence of twelve seconds \citep{aia_overview}.

    By examining the solar EUV images from the A94, A335, and A171 channels of AIA (see Table~\ref{table:eve_lines} for the dominant flare line in each) we can determine the dynamics of the C8.8 flare and the physical nature of the EUV late phase. A movie made using the AIA images from this flare is available online. Select images from these AIA channels as well as running difference images from A171 are shown in Figure~\ref{fig:aia_eve}. These images are cutouts of of Active Region 11069 and cover the eruption, gradual phase, and EUV late phase of the flare from 11:47 UT to 14:25 UT. The GOES SXR and EVE lightcurves from Figure~\ref{fig:sxr_eve_flare_phases} are shown for reference with the time of each AIA image marked.

    %Figure: Overview of flare with both AIA and EVE

    \subsection{Preflare Configuration and Flare Onset}

\begin{figure}[htbp]
    \begin{center}
    \includegraphics[width=\columnwidth]{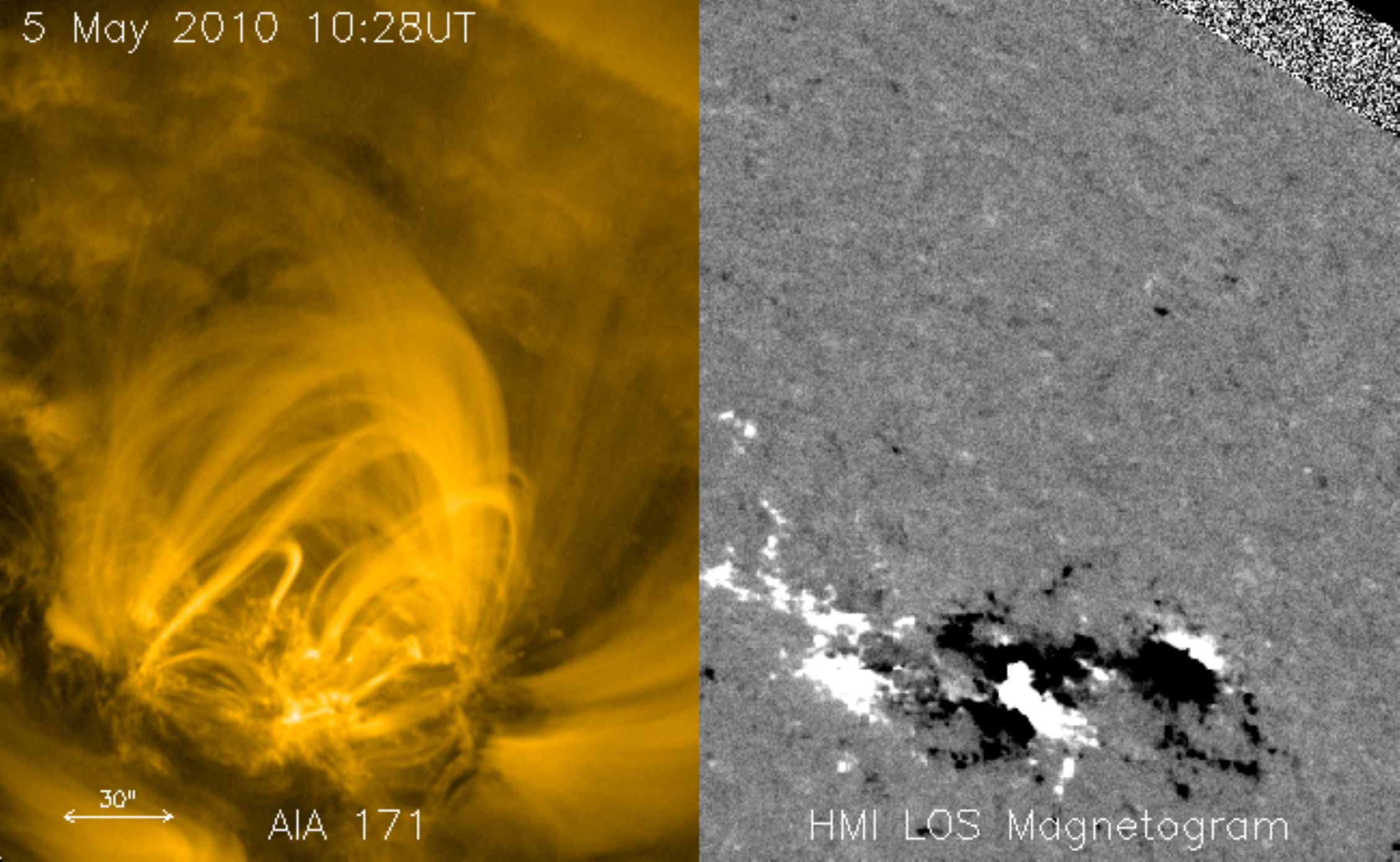}
    \end{center}
    \caption{Configuration of Active Region 11069 prior to the C8.8 flare on 2010 May 5.  The AIA A171 image (left) shows the coronal structure of the active consisting of a set of inner loops, two side lobes, and a set of overlying loops.  The HMI line-of-sight magnetogram (right) shows an almost linear quadrupolar configuration.}
    \label{fig:late_phase_config}
    \end{figure}

\begin{figure}[htbp]
    \begin{center}
    \includegraphics[width=\columnwidth]{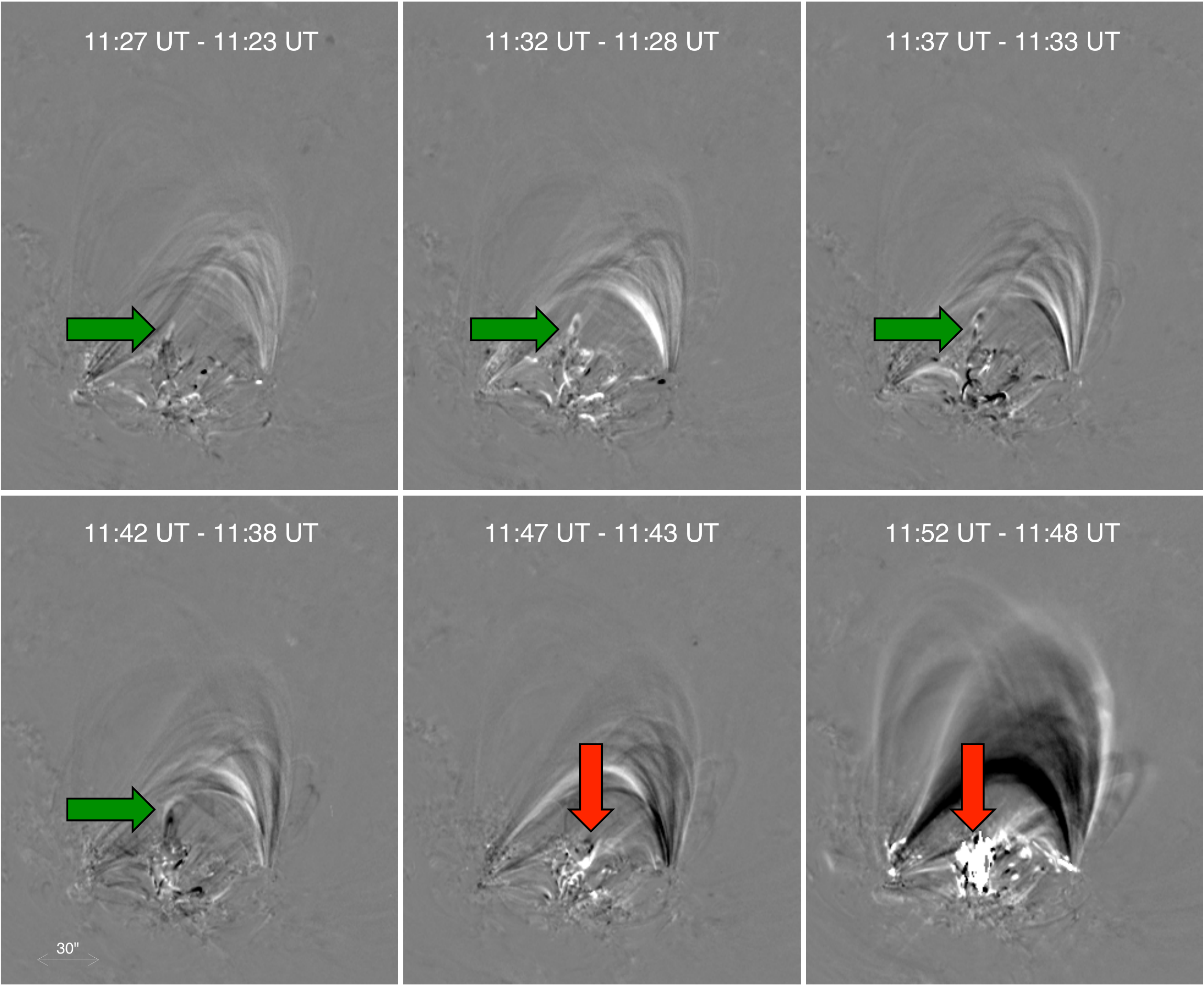}
    \end{center}
    \caption{Running difference images from AIA 171 show rising loops in the core of the active region (green arrows) prior to onset of the C8.8 flare.  The C8.8 flare occurs in the active region core (red arrows). Simultaneous to the onset of the C8.8 flare, the overlying loops begin to expand outward.}
    \label{fig:aia_rising_loops}
    \end{figure}

    The C8.8 flare on 2010 May 5 occurred in Active Region 11069, a magnetically complex region located at N42W35. The location of this flare near the western limb of the solar disk, but not over the limb, provides a viewing angle that allows see us to observe both the loop structure in AIA images and the magnetic field configuration from line-of-sight magnetograms from HMI.  Looking at a pre-flare AIA 171 image from 10:28 UT  (Figure~\ref{fig:late_phase_config}, left) and a co-temporal HMI line-of-sight magnetogram (Figure~\ref{fig:late_phase_config}, right), the active region conforms to a quadrupolar with an almost linear alignment of the different flux systems. Prior to the C8.8 flare, there was ongoing evolution of the active region and small brightenings and loop movements. Upward motions are visible as early as 11:16 UT as shown in Figure~\ref{fig:aia_rising_loops}, but rapid expansion does not commence until the flare begins at 11:49 UT. The events shown in Figure~\ref{fig:aia_rising_loops} occur prior to and during times marked 1 and 2 in Figure~\ref{fig:aia_eve}. The rising of the inner loops observed prior to the C8.8 flare are a result of breakout reconnection at the magnetic null point in the schematic model (Figure~\ref{fig:cartoon}).

    % Figure: Late phase configuration

    % Figure: AIA Flare Trigger

    \subsection{Impulsive and gradual phases}

    After the onset of the C8.8 flare at 11:49 UT, the main phase of the flare occurs (corresponding to times marked 2--4 in Figure~\ref{fig:aia_eve}) .  This main phase includes the impulsive phase and the gradual phase. The impulsive phase of flare consists mostly of non-thermal emission while the gradual phase consists of thermal emission from the heating and subsequent cooling of recently reconnected post-flare loops and is seen in both GOES SXR and coronal EUV lines.

\begin{figure}[htbp]
    \begin{center}
    \includegraphics[width=\columnwidth]{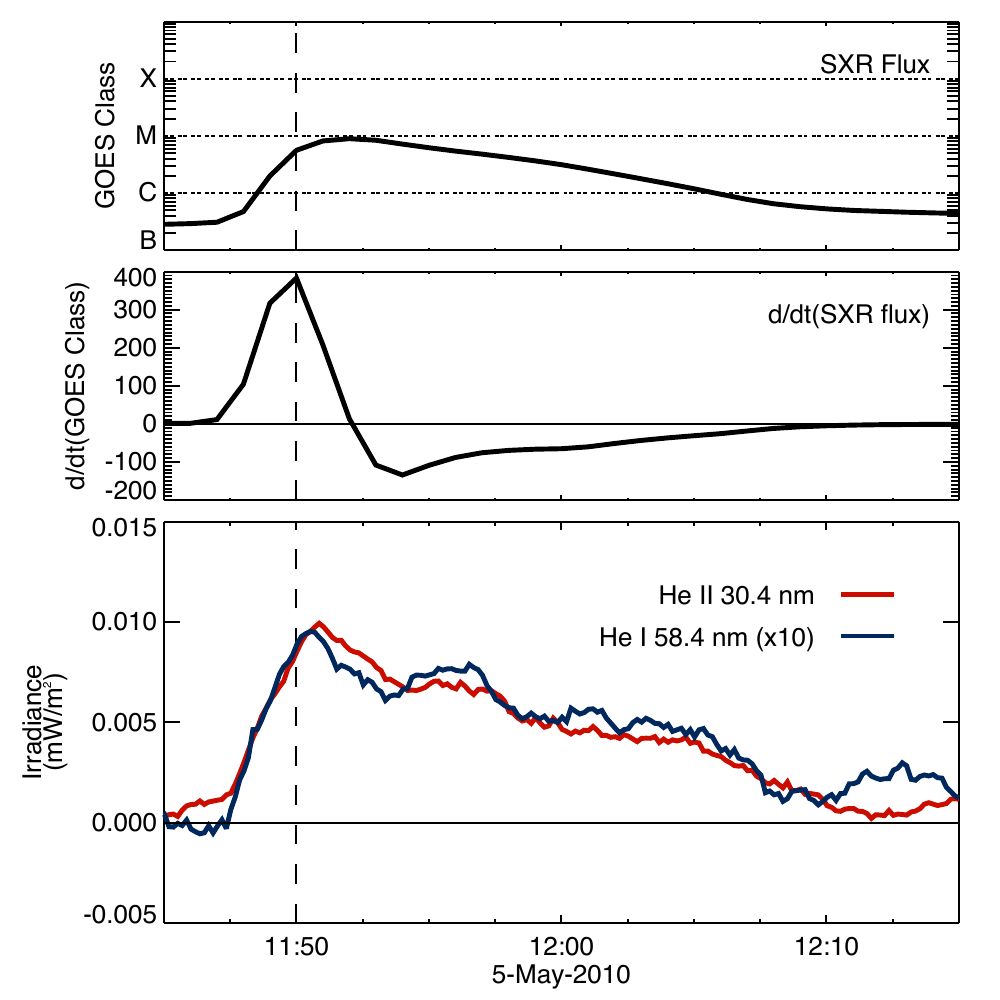}
    \end{center}
    \caption{Impulsive phase of the 2010 May 5 C8.8 flare is seen in transition region emissions such as He {\sc ii} 30.4 nm and He {\i} 58.4 nm (bottom panel).  These lines peak prior to the GOES soft X-ray (top panel) and follow the Neupert effect, which implies that the impulsive phase should peak at the maximum of the time derivation of the GOES SXR (middle panel). The dashed line shows the timing of the maximum of the GOES SXR time derivative.}
    \label{fig:eve_impulsive_phase}
    \end{figure}

    The impulsive phase is usually observed in non-thermal hard X-ray emission (HXR) by such satellites as RHESSI \citep{rhessi} but can also be seen in transition region emission lines, which EVE measures. Although there is no RHESSI data for this flare, the transition region emissions from EVE allow us to explore the relationship between the impulsive phase and the rest of the flare. The impulsive phase peaks prior to the peak in the GOES SXR, which is representative of the gradual phase.  This relationship between the impulsive and gradual phases is known as the Neupert effect \citep{neupert_neupert_effect}: the time-integrated hard X-ray emission ($F_{HXR}$) is proportional to the soft X-ray emission ($F_{SXR}$):
    \begin{equation}
    \int_{0}^{t}{F_{HXR}\left(t'\right)} dt' \propto F_{SXR}\left(t\right)
    \end{equation}
As shown in Figure~\ref{fig:eve_impulsive_phase}, the time derivative of the GOES SXR also correlates reasonably well with the transition region emissions, such as He {\sc ii} at 30.4 nm and He {\sc i} at 58.4 nm: the transition region emissions peak at the same time as the time derivative of the GOES SXR.  Both the EVE lines and the GOES SXR time derivative peak prior to the GOES SXR.

    % Figure: EVE Impulsive Phase

\begin{figure}[htbp]
    \begin{center}
    \includegraphics[width=\columnwidth]{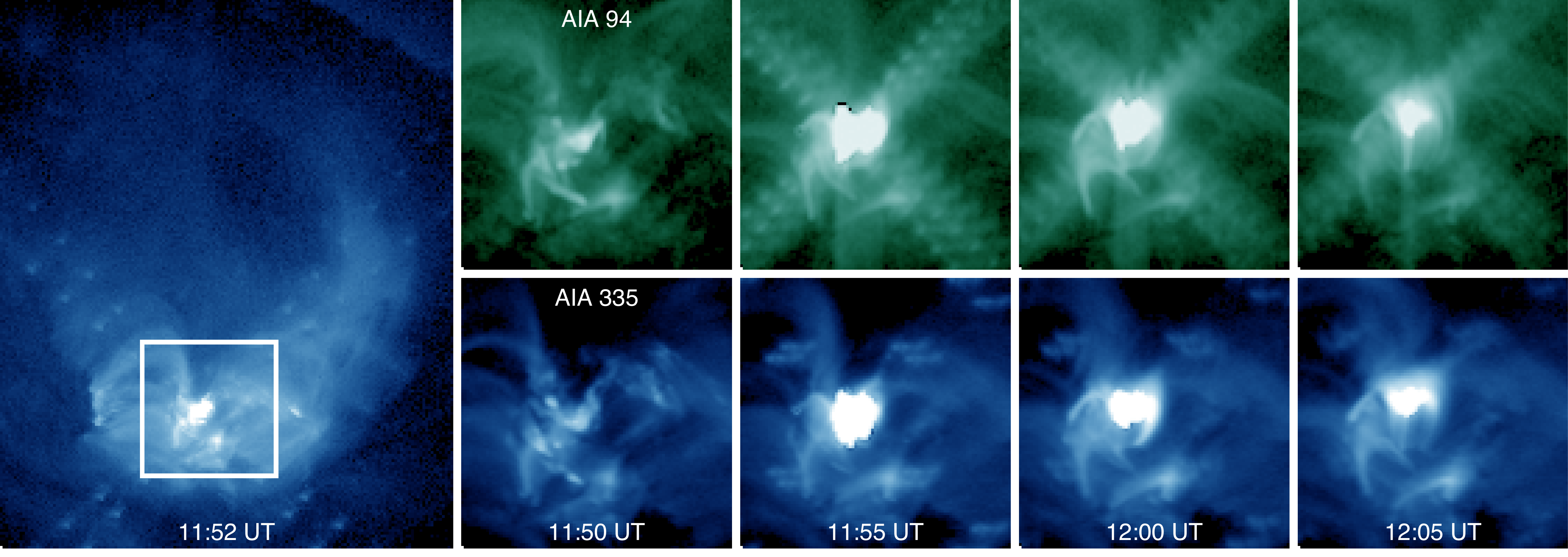}
    \end{center}
    \caption{Gradual phase of the 2010 May 5 C8.8 flare as seen by AIA A94 and A335 channels.  The figure on the left shows the entire active region with cutout region in white.  On the right are a series of images from A94 (top row) and AIA A335 (bottom row).  Time moves from left to right.  While parts of the images saturated and show diffraction during the actual flare, it is possible to observe the cusp-shape of the post-flare loops while they cool. }
    \label{fig:aia_gradual_phase_details}
    \end{figure}

    Because the AIA images saturated at the onset of the flare, it is impossible to know exactly what part of the coronal loop is brightening during the flare.  By looking at the flaring region immediately after the peak of the flare (Figure~\ref{fig:aia_gradual_phase_details}), it is possible to discern that the bright emission in the A94 and A335 channels  on AIA is coming from the tops of a small arcade of cusp-shaped loops. These cusp-shaped loops are indicative of recently reconnected coronal loops and that AIA in these channels is observing the gradual phase of the flare with most of the emission coming from the bright loop tops.

    % Figure: AIA Gradual Phase Details

\begin{figure}[htbp]
    \begin{centering}
    \includegraphics[width=\columnwidth]{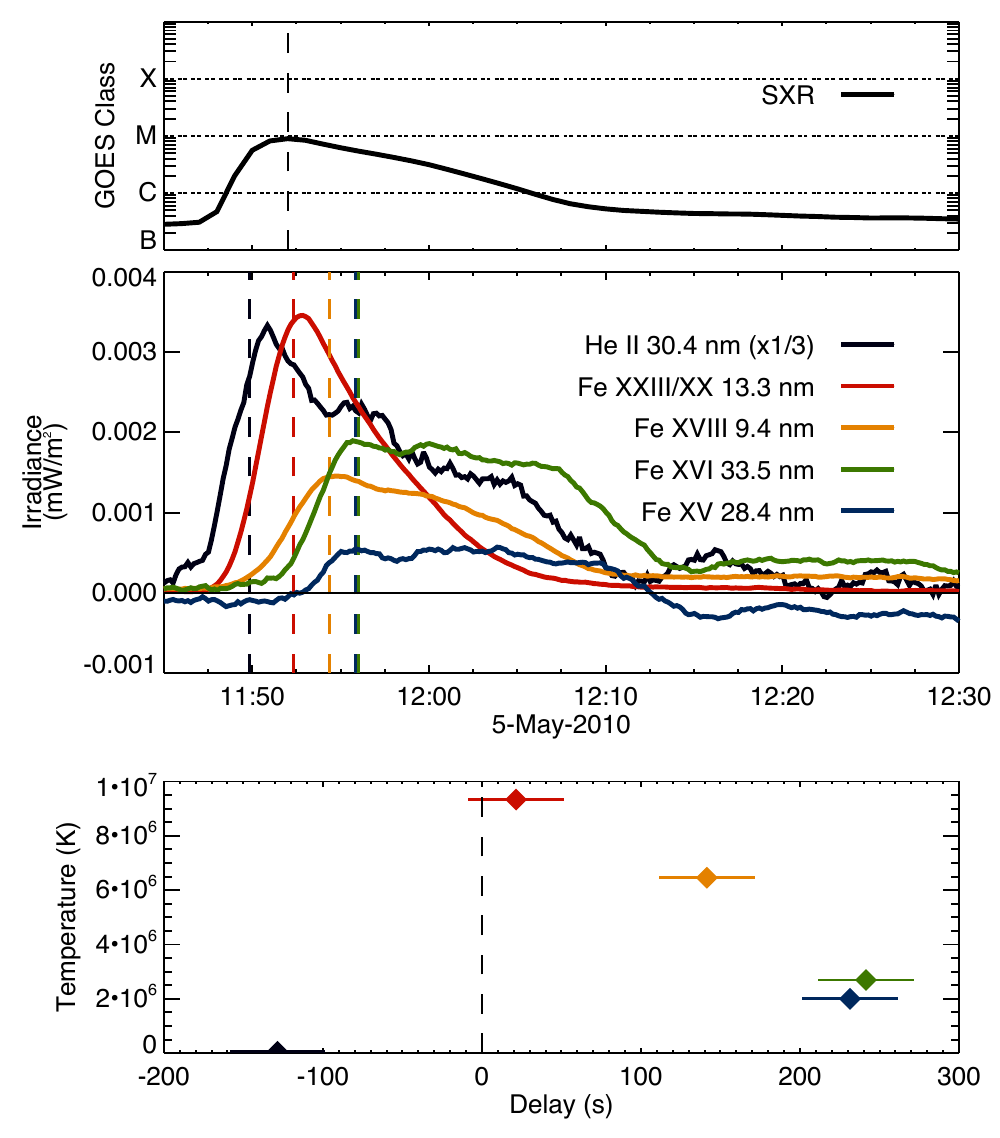}
    \end{centering}
    \caption{Gradual phase of the 2010 May 5 C8.8 flare as observed by EVE. The top panel shows the GOES SXR with the peak denoted by a dashed line.  The middle panel shows the lightcurves from EVE for four coronal lines and the He {\sc ii} line at 30.4 nm.  The timing of the peak irradiance for each line is denoted by a dashed line in the corresponding color.  The bottom panel shows delay between the peak of each EVE line and the peak of GOES SXR versus temperature of the line.  The He {\sc ii} line peaks prior to GOES as is expected for an impulsive phase emission with the progressively cooler coronal lines peaking later.}
    \label{fig:eve_gradual_phase}
    \end{figure}

    The hotter coronal emissions observed by EVE further confirm that the bright emissions in the AIA A94 and A335 channels are the gradual phase of the flare. For this flare, the gradual phase is seen by EVE in the hot coronal lines from the Fe {\sc xxiii}/{\sc xx} blend at 10 MK down to the Fe {\sc xv} line at 2 MK (Figure~\ref{fig:eve_gradual_phase}). The peak time for each of these lines relative to the GOES SXR peak in shown Figure~\ref{fig:eve_gradual_phase}. The progressively cooler coronal lines peak later. These emissions are coming from plasma that is undergoing conductive and radiative cooling as expected for flare loops. The flare loops cool from ~10 MK to 2 MK in about 250 seconds which is consistent with previous work by \citet{cargill_rad_cooling}.

    % Figure: EVE Gradual Phase

    \subsection{Eruption, Formation of a CME, and Evidence of Breakout Reconnection}
    \label{sec:eruption_breakout}

    Although eruption is initiated by breakout reconnection before the flare occurs, the eruption is greatly accelerated by the flare reconnection, as shown in Figure~\ref{fig:aia_eve} (11:47 UT--11:59 UT).  The initial brightening of the flare is localized to the inner loop region of the quadrupole.  In the eruption, some loops are pushed aside while others open forming a CME that was observed by {\it SOHO}'s Large Angle and Spectrometric Coronagraph (LASCO). According to the CDAW LASCO CME catalog, the CME has a narrow angular extent ($7^\circ$) and is extremely faint (only detected in inner C2 channel).  This CME is clearly not geoeffective and it could be argued that it is not a ``real'' CME.  The presence of this faint CME in the LASCO data, however, confirms an eruption in the AIA images, which is important for understanding the origin of the EUV late phase.
    
    In Figure~\ref{fig:cartoon}, the CME is shown in 2D as a plasmoid (purple lines).  Given that the plasmoid originates from the inner loops and the small size of the active region, it is not surprising that the observed CME is small and faint.

\begin{figure}[htbp]
    \begin{center}
    \includegraphics[width=\columnwidth]{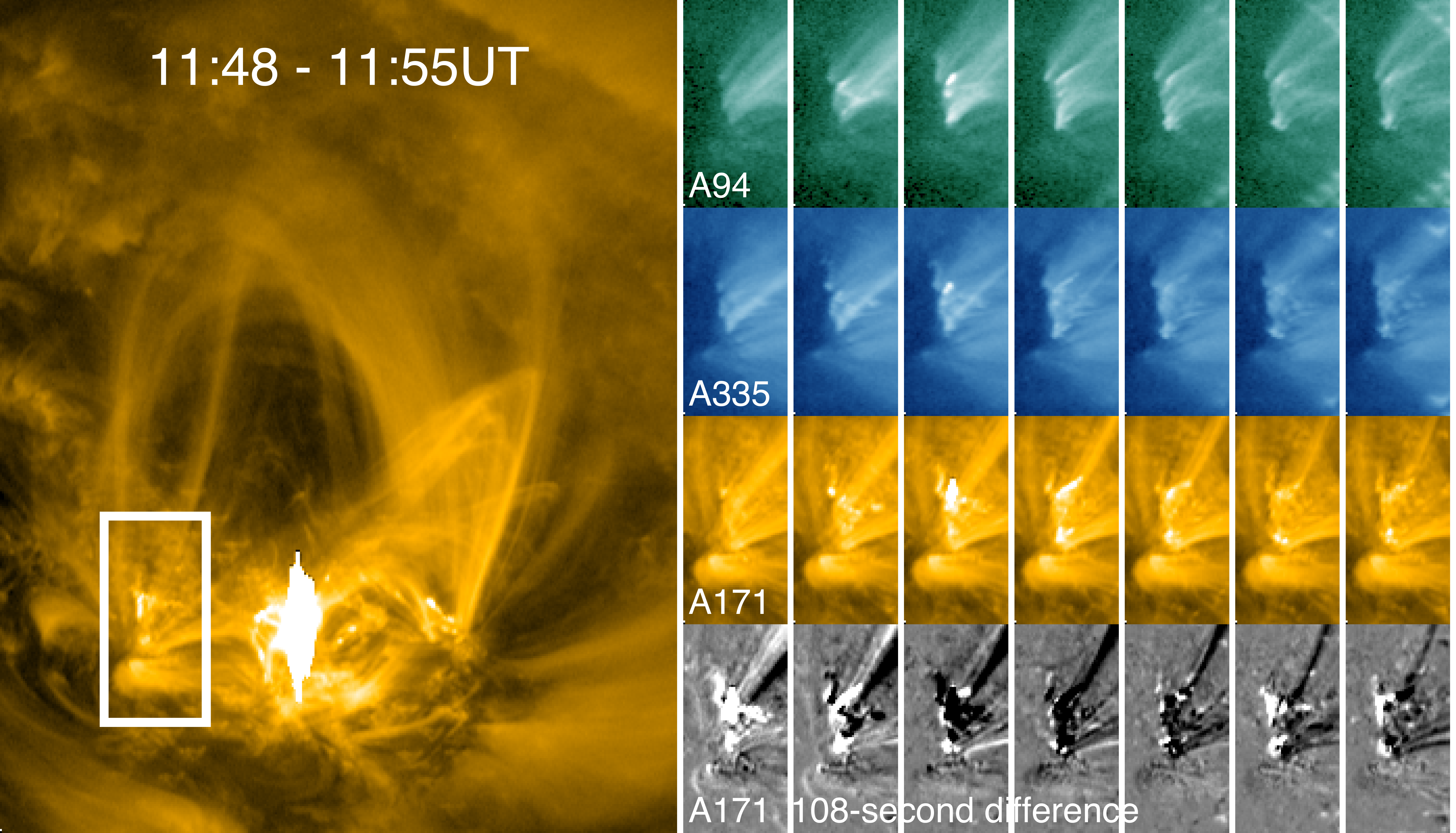}
    \end{center}
    \caption{Evidence of breakout reconnection.  The image on the left shows the entire active region and highlights the cutout region.  On the right, there is a series of cutouts of the footpoint region for AIA A94, A335, and A171 channels as well as running difference images from A171 (one image every minute with time moves from left to right).  Looking at the footpoints of the outer loops during the eruption, we can see intense brightening of the footpoints in all three AIA channels.  The running difference images show a slight movement of the brightenings outward. }
    \label{fig:aia_breakout}
    \end{figure}

	Although breakout reconnection is considerably less energetic than
flare reconnection, evidence for energy release can be found.
Figure~\ref{fig:aia_breakout} shows brightening at the footpoints of
the outermost side lobe field loops that are shown schematically in
green in Figure~\ref{fig:cartoon}.  Recall that these loops are
created when inner (red) and outer (blue) field lines reconnect at
the null point. As is expected from reconnecting successive loops,
there is a slight apparent movement of the bright footpoints
outward.

    % Figure: Breakout reconnection

    \subsection{EUV Late Phase and Active Region Recovery}

    By 12:25 UT, the inner coronal loops, which were heated during the main phase of the flare, have cooled (see Figure~\ref{fig:aia_eve}). Additional emission from loops overlying the active region begins to appear around 13:00 UT.  It is seen in the AIA A335 channel by 13:05 UT. By 13:45 UT, loops are visible in the A171 channel. This emission above the core of the active region forms the EUV late phase that is observed by EVE (see Figure~\ref{fig:sxr_eve_flare_phases} and Figure~\ref{fig:aia_eve}). There are three points to consider when discussing the EUV late phase emission: Where does this emission originate?; What is its evolution in the active region?; How is the EUV late phase different from normal active region evolution?

\begin{figure}[htbp]
    \begin{center}
    \includegraphics[width=\columnwidth]{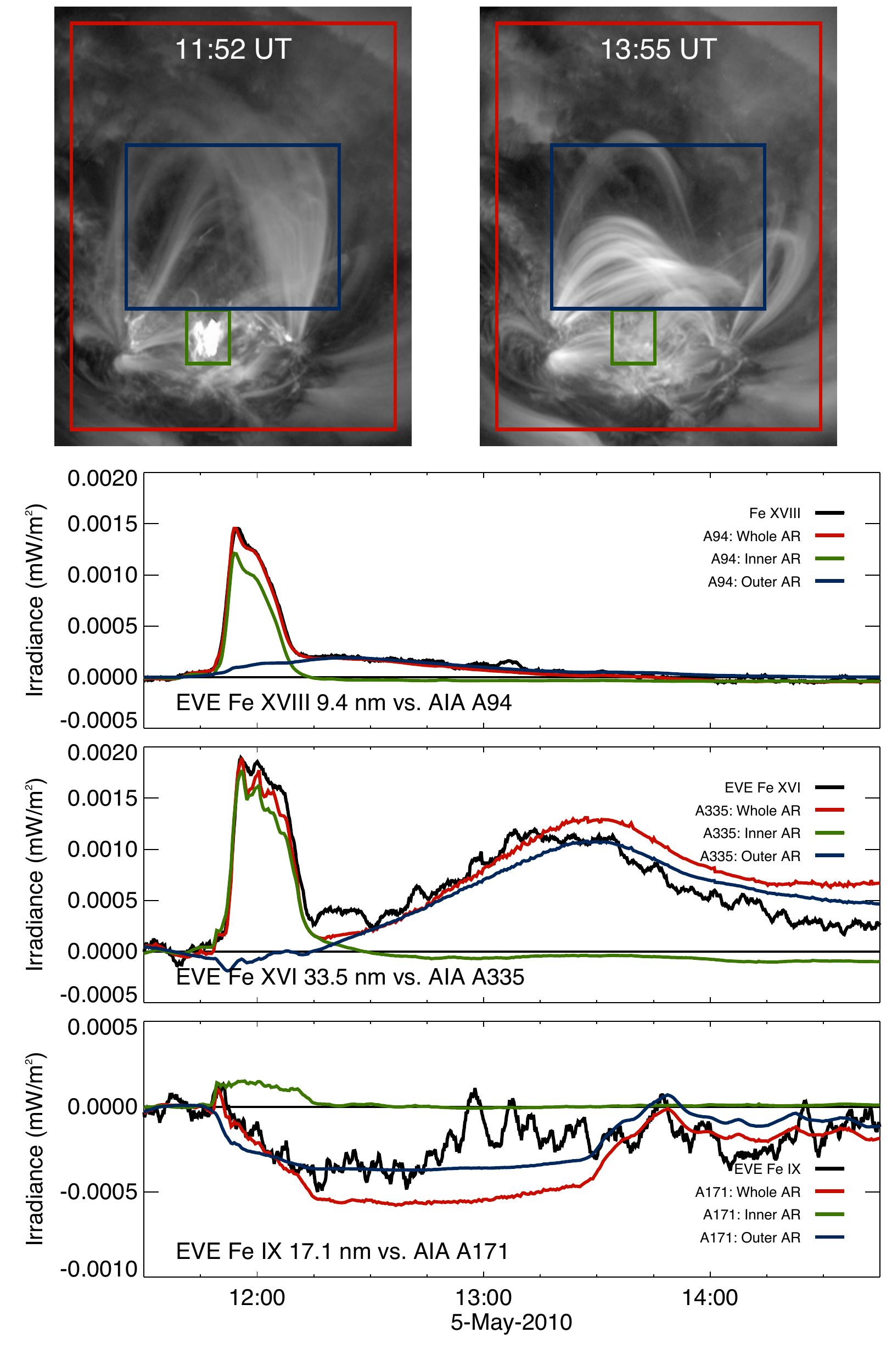}
    \end{center}
    \caption{Comparison of AIA and EVE lightcurves.  Three sets of lighcurves were made from AIA images by summing the count rate (DN/s) for every pixel in a particular region.  The regions used are shown at the top.  The red region outlines the entire active region; the green the core of the active region; the blue the overlying loops. The bottom of the figure compares these AIA lightcuves with EVE lightcurves for the dominant line in each AIA bandpass.  The emissions during the main phase of the flare come solely from the core of the active region while the EUV late phase originates in the overlying loops. While both the main and EUV late phase are visible in individual AIA A171 images, the lightcurve of the entire active region is dominated by coronal dimming.}
    \label{fig:aia_eve_compare}
    \end{figure}

    To determine where the EUV late phase emission originates, we compare the AIA light curves from different parts of the active region with EVE light curves of the dominant emission line in AIA bandpass (Figure~\ref{fig:aia_eve_compare}). The AIA lightcurves are calculated by summing the count rate (DN/s) for every pixel in each image for three different regions (red is the whole active region, green is the core of the active region, blue is the overlying loops believed to form the EUV late phase) and subtracting a pre-flare value. The pre-flare value is determined using the same time range that is used to determine the pre-flare spectrum for the EVE data. A scale factor is applied to allow comparison between the uncalibrated AIA count rate and EVE irradiance lightcurves.  This scale factor is the ratio of the peak of the EVE lightcurve to the peak of the AIA lightcurve for the whole active region (red).  The same scale factor is applied for the three different extracted regions but varies between AIA channels.

    % Figure: AIA vs. EVE

    This comparison of the EVE and AIA lightcurves tells us two things. First, comparing the shape of the red curves, Figure~\ref{fig:aia_eve_compare} shows that the increase in EUV irradiances from EVE during this flare can be attributed wholly to the active region.  The first peak in the EVE lightcurves, associated with the main phase of the flare, corresponds to the changes in the inner core of the active region (Figure~\ref{fig:aia_eve_compare}, green curves). Although the core of the active region brightens in the AIA 171 channel, the dimming seen above the core of the active region dominates the lightcurve of the whole region. Second, the core of the active region does not contribute to the EUV late phase. The EUV late phase originates solely in the outer active region (Figure~\ref{fig:aia_eve_compare}, blue curves).  While all three channels show this emission, it is best observed in the AIA A335 channel. The peak in the EUV late phase is first seen in the A94 channel at 12:22 UT. A94 is dominated by a hot (10 MK) coronal line.  It is then seen in the cooler A335 channel (13:27 UT, 2-3 MK) and final seen in the A171 channel (13:49 UT, 1 MK).  This progression of the EUV late phase from hot to warm to cool coronal lines is evidence that something heated the plasma in the outer loops, which then radiatively cooled.  Because these loops appear to be newly formed, we believe some form of reconnection is responsible for the heating of the EUV late phase.
   
\begin{figure}[htbp]
    \begin{center}
    \includegraphics[width=\columnwidth]{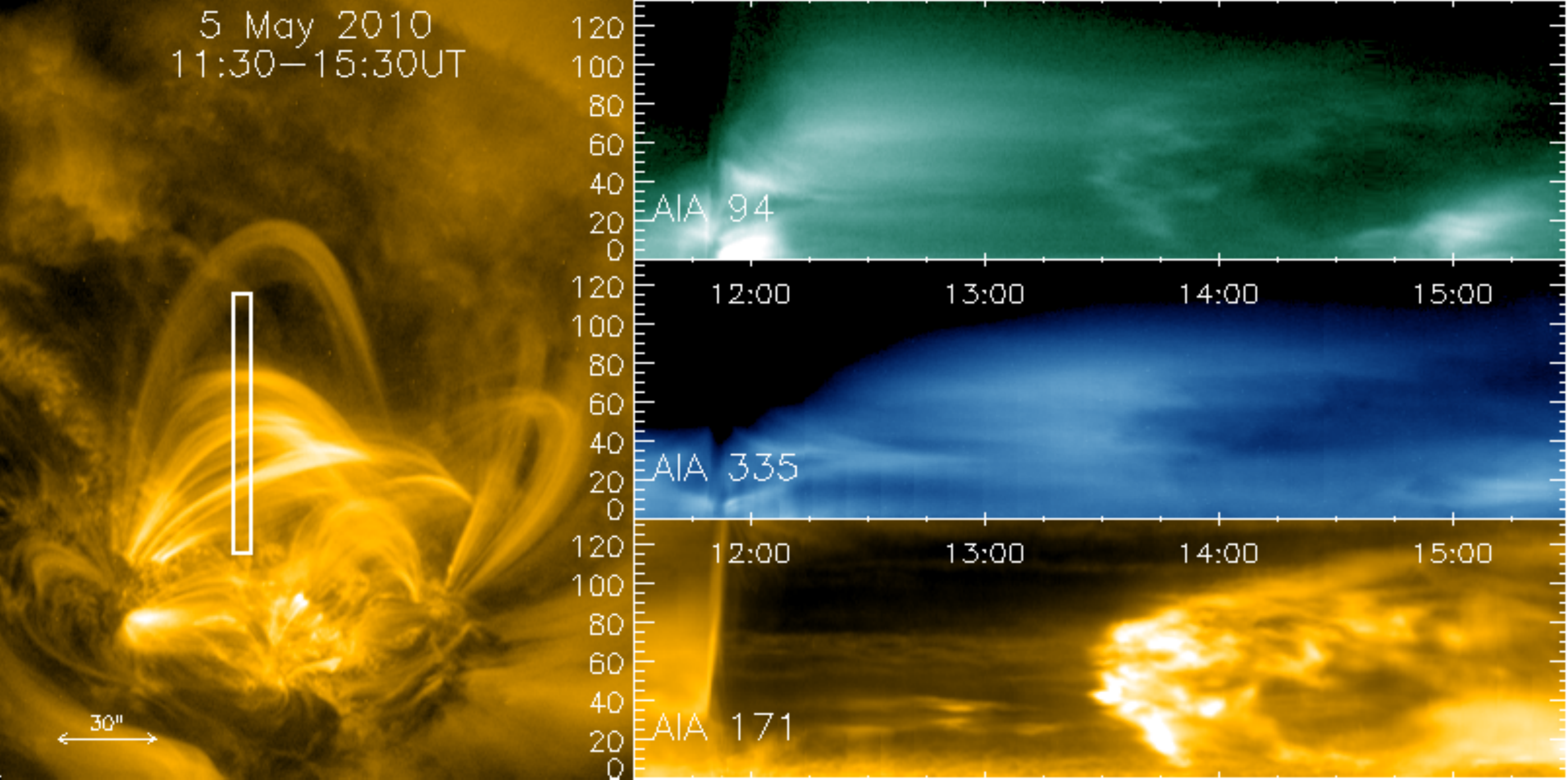}
    \end{center}
    \caption{Time-distance analysis of the EUV late phase.  By looking at a slice (identified on the left) of the overlying coronal loops as a function of time (right panels) for three AIA channels (A94, A335, and A171), we understand both the spatial, temporal, and thermal evolution of the coronal loops EUV late phase. The loops first appear in the hottest channel (A94) and last in the coolest channel (A171).  The loops also first appear at a lower height and then successively higher.}
    \label{fig:aia_late_phase_evolution}
    \end{figure}

    The behavior of the coronal loops in the outer active region is more clearly seen in Figure~\ref{fig:aia_late_phase_evolution}, which shows the time-distance evolution of EUV late phase coronal loops. By looking at a slice of the AIA images as a function time we can examine how the outer coronal loops evolve throughout the flare. The first thing to notice is the evacuation of the corona just prior to 12:00 UT.  This is most evident in the A171 channel and is a result of the eruption and launch of a CME (Section~\ref{sec:eruption_breakout}). Shortly afterwards, we start to see enhanced emission first in the A94 channel, followed by brightening in the A335 and A171 channel.  This is consistent with radiative cooling of coronal loops from hot emission in the A94 bandpass to warm emission in A335 and cool emission in A171, which was also observed in Figure~\ref{fig:aia_eve_compare}.

    % Figure: Time-distance Plot for Late Phase

	The EUV late phase is the result of organized, successive reconnection and is different than typical active region evolution.  The upward trend of the bright
emission in Figure~\ref{fig:aia_late_phase_evolution} tells
us that the emission is coming from higher loops later in the EUV
late phase.  This is expected based on Figure~\ref{fig:cartoon}d. As
field lines reconnect high in the corona and retract, they
accumulate on top of previously reconnected field lines and
therefore do not retract as far.

The final thing to consider is how the EUV late phase differs from the main phase of the flare. As discussed above, the EUV late phase originates in the outer part of the active region while the main phase of the flare occurs in the core of the active region.  As a result, the EUV late phase is associated with longer and higher loops than the main phase.  

Furthermore, when looking at the irradiance lightcurves from EVE, the EUV late phase lasts longer and there is a greater delay in the peaks of the irradiance between Fe {\sc xvii} 9.4 nm to Fe {\sc xvi} 33.5 nm. For the main phase, this delay is about 2 minutes.  For the EUV late phase, it is 65 minutes. This is not the radiative cooling time of individual coronal loops (which is on the order of a few hundred seconds). It is the aggregated effect of many loops being heated at different times.

% ===============================
% Section 4: EBTEL Modeling
% ===============================

    \section{Modeling the Radiative Output Using EBTEL}
    \label{sec:ebtel}

    To understand why the EUV late phase is observed by EVE in the Fe {\sc xvi} line without no corresponding emission in hotter lines, we use the Enthalpy-Based Thermal Evolution of Loops (EBTEL) model \citep{klimchuk_ebtel} to estimate the peak temperatures and densities for the gradual phase and the EUV late phase. The EBTEL code was developed as a ``highly efficient model ... to describe the evolution of the average temperature, pressure, and density along a [single] coronal strand'' \citep{klimchuk_ebtel}. Combined with the CHIANTI atomic database \citep{chianti1, chianti2}, EBTEL can be used to predict lightcurves for specific emission lines.  EBTEL is considered to be a ``0D'' model as it calculates field-aligned averages instead of solving the hydrodynamical equations for multiple points along the coronal strand.  While this simplifies the physics, the EBTEL code is not computationally expensive. This enables us to develop a multiple-strand model of the flare based on the EBTEL code and determine the best-fit parameters to match the EVE data.

% Section 4.1: EBTEL model
    \subsection{EBTEL-based Flare Model and Model Parameterization}

    For a given heating rate profile and loop length, the EBTEL code returns the average coronal temperature, pressure, and electron density as a function of time as well as a time-dependent differential emission measure (DEM) for a single coronal loop or loop strand. The DEM can then be combined with atomic data from CHIANTI to calculate the radiance for any coronal emission line as a function of time. We developed an EBTEL-based flare model that uses multiple EBTEL loops with variable loop lengths and heating rate profiles to calculate the theoretical lightcurves  for the Fe {\sc xx}/{\sc xxiii} blend at 13.3 nm (one of the hottest and brightest flare lines observed by EVE) and the Fe {\sc xvi} line at 33.5 nm (to capture the EUV late phase).  Using MPFIT \citep{mpfit}, a non-linear least square fitting routine in IDL, we can optimize these input parameters so that the model lightcurves match the EVE lightcurves for those two lines.

    There are several steps to calculate the model lightcurves from the output of the EBTEL code. First, running the EBTEL code returns a differential emission measure as a function of time and temperature, $DEM\left(t, T\right)$, for a given heating rate profile, $h\left(t\right)$, and loop half-length, $l$. In the EBTEL-based flare model, we heat each loop once using a triangular heating profile that can be parameterized as:
    \begin{equation}
    h\left(t\right) =
        \begin{cases}
        h_{bkgd} + \frac{2h_{0}}{dt}\left(t-t_{0}\right) & \text{if $t_{0}-\frac{dt}{2}<t<t_{0}$,}
        \\
        h_{bkgd} - \frac{2h_{0}}{dt}\left(t-t_{0}\right) & \text{if $t_{0} \le t < t_{0}+\frac{dt}{2}$.}
        \end{cases}
    \end{equation}
The background level of heating, $h_{bkgd}$, is fixed at $10^{-6}$ erg cm$^{-3}$ s $^{-1}$ as is the duration of heating (300 s). The time of the heating, $t_{0}$, and strength of heating, $h_{0}$, are free parameters.
    
    The DEM output of EBTEL can then be turned into line radiance, $I_{loop}\left(t\right)$, via:
    \begin{equation}
    I_{loop}\left(t\right) = \int_T{DEM\left(t, T\right)G\left(T\right)dT}
    \end{equation}
where $G\left(T\right)$ is the contribution function calculated from CHIANTI.

    Finally, the radiance contributions from individual EBTEL loops is summed and converted in irradiance:
    \begin{equation}
    \label{eqn:i_loop}
    E_{model}\left(t\right) =  E_{pre-flare} + C\sum_{i=1}^{N_{loops}}{A_{i} I_{loop,i}\left(t\right)}
    \end{equation}
where $C$ is a normalization factor, $N_{loops}$ is the number of EBTEL loops, and $A_{i}$ is the loop cross-section. This model irradiance, $E_{model}\left(t\right)$, can then be compared to the EVE lightcurves and is the quantity that is optimized using MPFIT.

A summary of the parameters in the EBTEL-based flare model is shown in Table~\ref{table:ebtel_params}.  The first parameter in our EBTEL-based flare model is the total number of EBTEL loops ($N_{loops}$).  Every time the flare model runs, it calls EBTEL $N_{loops}$ times. As a result, this number needs to be large enough to capture the evolution of the flare but small enough that optimizing the model inputs is feasible. For the results presented in this paper, we choose to model twenty-two loops.

    % Table: EBTEL parameters
    \begin{table}[htdp]
    \caption{Parameters for the EBTEL-based Flare Model}
    \begin{center}
    \begin{tabular}{rl}
    Global parameters: \\
    $N_{loops}$ & Number of EBTEL loops \\
    $C$ & Normalization factor (sr m$^{-2}$)\\
    $h_{bkgd}$ & Background heating rate (erg cm$^{-3}$ s $^{-1}$)\\
    For each EBTEL loop:\\
    $l$ & Loop half-length (cm) \\
    $t_{0}$ & Time of peak impulsive heating (s)\\
    $dt$ & Duration of heating (s)\\
    $h_{0}$ & Strength of heating (erg cm$^{-3}$ s $^{-1}$)\\
    $A$ & Loop cross-section (cm$^2$) \\
    For each EVE line:\\
    $E_{pre-flare}$ & Preflare irradiance (W m$^{-2}$)\\
    \end{tabular}
    \end{center}
    \label{table:ebtel_params}
    \end{table}

    The loop cross-section, $A$, is a multiplicative factor that includes both the physical cross-section of the coronal loop and number of coronal loops.  Each EBTEL strand we model could represent one large coronal loop or may represent many smaller coronal loops. By allowing the cross-section to be larger than physically reasonable, we can model hundreds of loops with only a few EBTEL runs.

    The loop half-length, $l$, is a free parameter that is constrained using the AIA observations.  From AIA observations, we are able to determine that the coronal loops in the gradual phase of the flare are about 30 arc-seconds ($\approx$20 Mm) long while the loops in the late phase are about 100 arc-seconds ($\approx$70 Mm) long.  These are the lengths in the plane of the sky so the actual loops could be longer.
    
    To convert the output of EBTEL to irradiance so that it can be compared with EVE observations, the EBTEL output is multiplied by      a normalization factor, $C$, which is the solid angle subtended by 1 m$^2$ at 1 AU.

    Some of the parameters are coupled: the number of EBTEL loops and loop cross-section.  The loop cross-section is related to the number of strands.  If we model more strands, each strand will have a smaller cross-section. We have chosen the number of loops to be the fewest that gives us reasonable results. The loop length is constrained to be realistic based on the AIA observations while the timing and strength of the heating is allowed to vary freely.

% Subsection 4.2: Model results/discussion
    \subsection{Results and Discussion}

\begin{figure}[htbp]
    \begin{center}
    \includegraphics[width=\columnwidth]{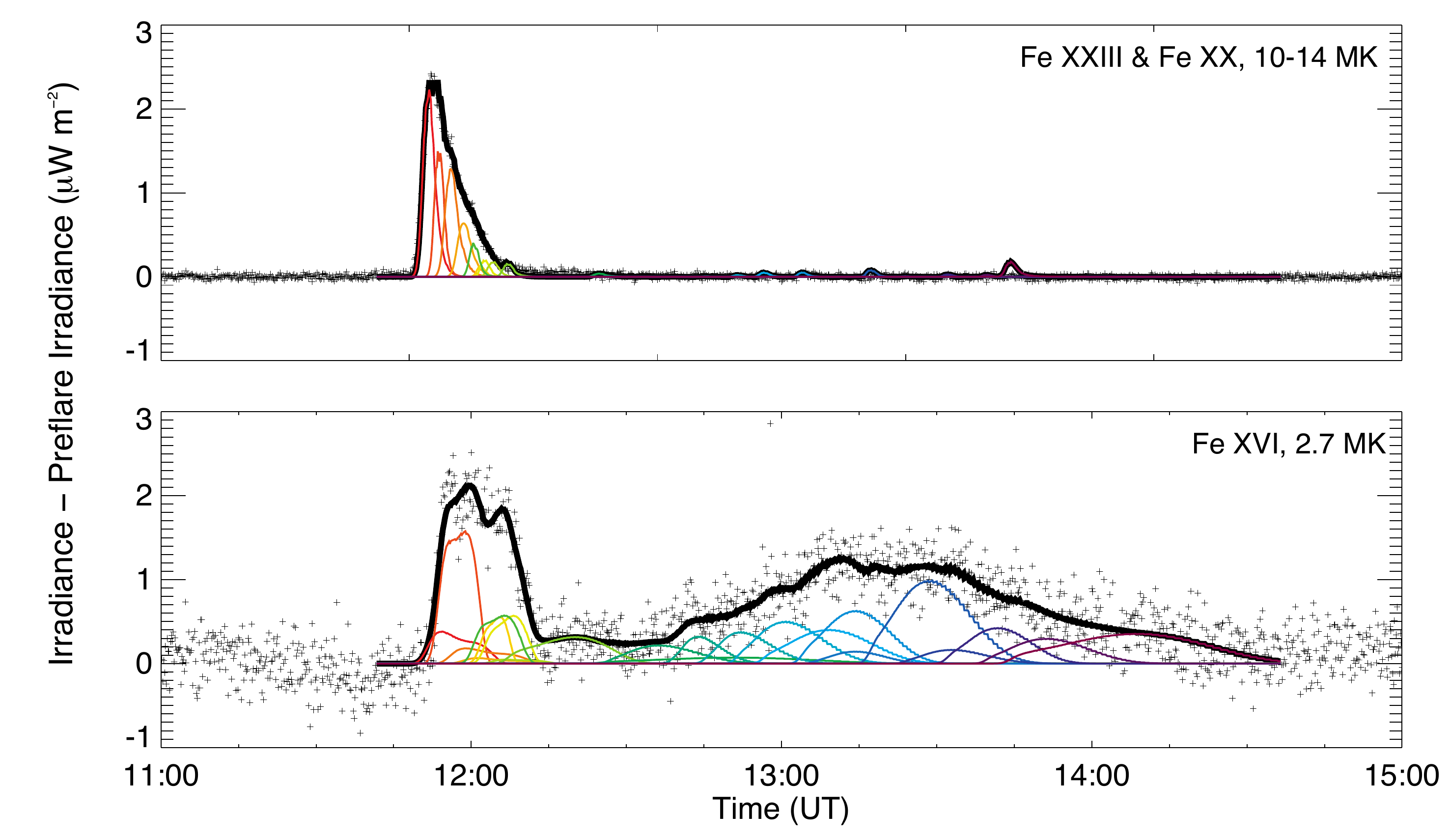}
    \end{center}
    \caption{Results of the EBTEL-based flare model for both the Fe {\sc xx}/{\sc xxiii} blended line (top) and the Fe {\sc xvi} line (bottom).  The pluses are the EVE observations with the pre-flare background irradiance subtracted off. The solid black line is the output from the model and agrees with the EVE lightcurves.  The individual colored lines are the contribution from each individual coronal loop strand.}
    \label{fig:ebtel_results}
    \end{figure}

    Using the EBTEL-based flare model described above with $N_{loops}=22$, we are able to determine the best-fit parameters to describe the 13.3 nm (Fe {\sc xx}/{\sc xxiii},  10-14 MK) and the 33.5 nm (Fe {\sc xvi}, 2.7 MK) lightcurves from EVE. The resulting theoretical and measured lightcurves are shown in Figure~\ref{fig:ebtel_results}. There is good agreement between the lightcurves measured by EVE and the theoretical lightcurves calculated from the EBTEL-based flare model. Both the gradual phase and EUV late phase must be modeled with many loops.  The EUV late phase, in particular, results from heating many loops consecutively through the duration of the EUV late phase.

    % FIgure: EBTEL results

\begin{figure}[htbp]
    \begin{center}
    \includegraphics[width=\columnwidth]{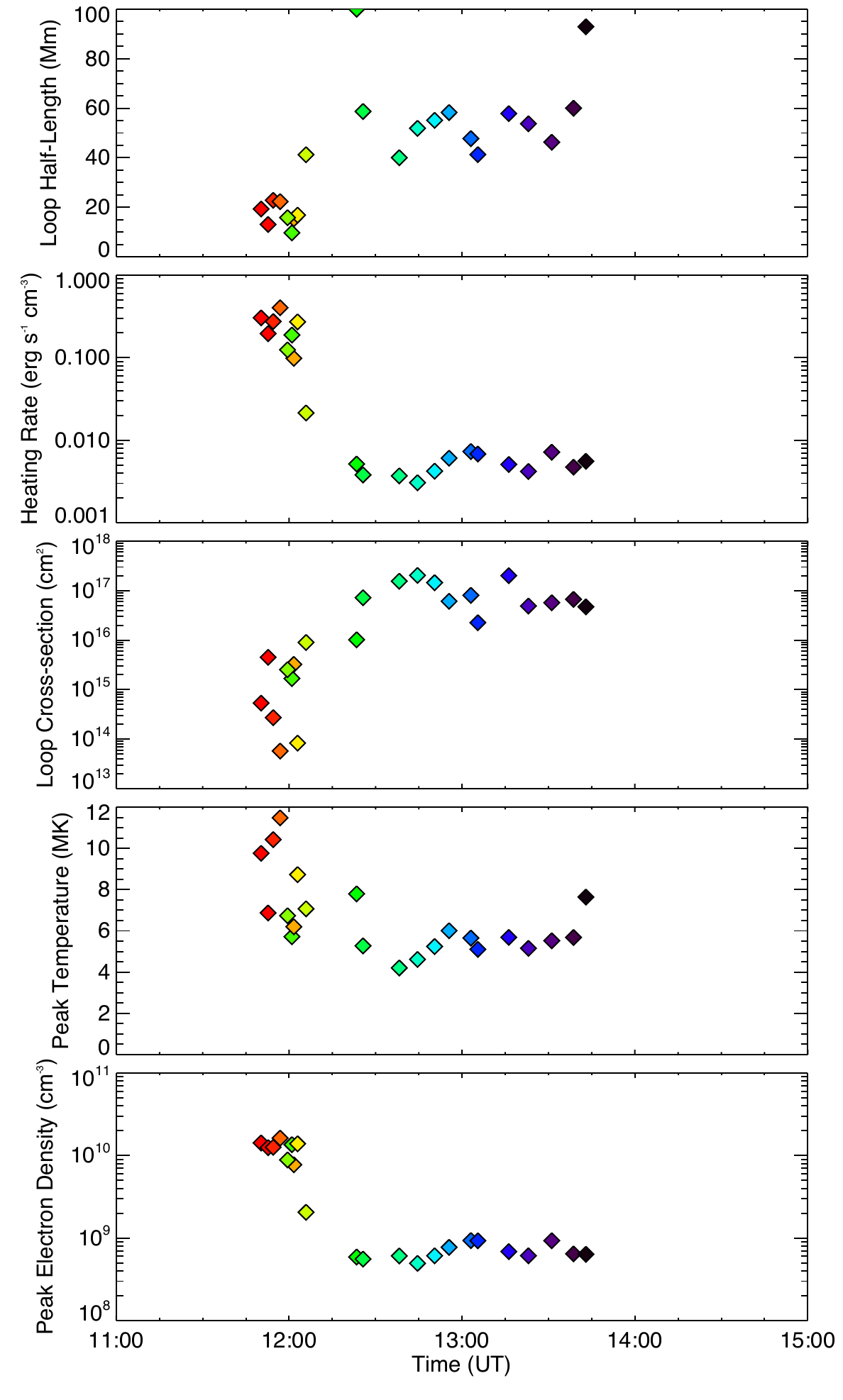}
    \end{center}
    \caption{Best-fit parameters of EBTEL-based flare model.  The colored diamonds are the parameter for each individual coronal loop strand and correspond to the lightcurves in the pervious image.  Each parameter is plotted as a function of time or when each loop was heated in the model.  AIA images give loop half-length estimates of 5-15 Mm for the main phase and 35-50 Mm for the EUV late phase.  These are plane-of-the-sky estimates, giving the lower boundary for the loop half-length.}
    \label{fig:ebtel_parameters}
    \end{figure}

    The best-fit model parameters are shown in Figure~\ref{fig:ebtel_parameters} along with the peak temperature and electron density for each EBTEL loop. The heating profile from this model clearly shows that there are two separate phases of heating.  The first heating event involves the gradual phase. A large of amount of heating occurs within a short time.  As the loops cool, emissions are seen in progressively cooler lines from 10 to 1 million K.  The second phase is very different.  A large number of coronal loops are heated only slightly.  The plasma in each loop, instead of reaching 10 million K, reaches 5 million K.  The heating is also spread out over an hour generating the long secondary flare emission profile.  This model is consistent with AIA observations that the secondary flare emission comes from the active region restructuring itself after a flare.

% FIgure: EBTEL parameters

% ===============================
% Section 5: Discussion and Conclusions
% ===============================

    \section{Discussion and Conclusions}
    \label{sec:discussion}

    Here, we have examined one EUV late phase flare from 5 May 2010 in detail to understand the origin and nature of the EUV late phase. The EUV late phase is a second increase in the irradiance of warm coronal lines (2-3 MK) such as the Fe XVI 33.5 nm line after the gradual phase of a flare without a corresponding increase in hotter emission lines such at Fe XVIII 9.4 nm. From examining AIA images, the EUV late phase is clearly associated with the C8.8 flare but is both temporally and spatially separate. The EUV late phase originates in the reforming of overlying loops that are opened when a small CME erupts during the flare.

    Using EBTEL to model the EVE irradiance lightcurves, we have shown that the enhancement in Fe XVI 33.5 nm but not Fe XVIII 9.4 nm during the EUV late phase is possible by heating the coronal loops less that during the main phase of the C8.8 flare. The lower heating rate results in a lower peak temperature of the loop and a lower peak electron density, reducing the emission at the hotter emission lines.

    The model we propose here is similar to CME initiation model in \citet{ben_cme_model}, based breakout reconnection.  While the CME model focuses on how the CME erupts, our model looks at how the corona  reconfigures itself post-eruption.   Classical post-flare loops, which form the gradual phase of the flare, are a result of internal reconnection within an inner magnetic flux system (red in Figure~\ref{fig:cartoon}) and are associated with the early rapid acceleration of the CME.  The late phase involves a completely different set of loops that result from internal reconnection within a topologically different flux system (blue in Figure~\ref{fig:cartoon}).  As this much later reconnection involves weaker fields and produces longer loops, the heating rate per unit length is lower, and the EUV late phase is therefore best observed in warm coronal lines.  Note that the late phase can only occur if all of the inner (red) flux reconnects first and moves out of the way, so not all breakout eruptions are expected to have a late phase.  A possible different scenario is described at the end of Section~\ref{sec:model}.
    
    It is unlikely that EUV late phase flares are newly observed phenomenon.  The extended wavelength coverage, higher spatial resolution, and higher temporal cadence of AIA in addition to the irradiance measurements from EVE, however, make it possible to study these types of events in detail for the first time.

% ===============================
% Acknowledgements
% ===============================

    \acknowledgments
This research was supported by NASA contract NAS5-02140 to the University of Colorado. The work of JAK was supported by the NASA Supporting Research and Technology program.  SDO data courtesy of NASA/SDO and the AIA, EVE, and HMI science teams. This CME catalog is generated and maintained at the CDAW Data Center by NASA and The Catholic University of America in cooperation with the Naval Research Laboratory. SOHO is a project of international cooperation between ESA and NASA.

    {\it Facilities:} \facility{SDO (AIA, EVE, HMI)}
\clearpage
% ===============================
% References
% ===============================

\end{document}